\RequirePackage{fix-cm}
\documentclass[smallextended]{svjour3}       % onecolumn (second format)
\smartqed  % flush right qed marks, e.g. at end of proof
\usepackage{graphicx}
\usepackage{mathptmx}      % use Times fonts if available on your TeX system
\usepackage{tikz}
\usetikzlibrary{calc,patterns,decorations.pathmorphing,decorations.markings}

\usepackage{hyperref}
\usepackage{amsmath}
\usepackage[numbers]{natbib}

\usepackage{amssymb}

\newcommand{\Id}{\mathbb{I}}

\newcommand{\R}{\mathbb{R}}

\newcommand{\mat}[1]{\mathbf{#1}}
\newcommand{\dmat}[1]{\dot{\mat{#1}}}

\begin{document}

\title{A simplified model for the forced libration of icy moons with subsurface oceans%\thanks{Grants or other notes
}
\subtitle{Application to Enceladus and Mimas}


\author{Yeva Gevorgyan
}


\institute{Yeva Gevorgyan \at
              CEMSE Division, King Abdullah University of Science and Technology, Thuwal 23955-6900, Saudi Arabia. \\
              \email{yeva.gevorgyan@kaust.edu.sa}           }

\date{Received: 18 June 2025 / Revised: 6 November 2025 / Accepted: 8 November 2025}

\maketitle

\begin{abstract}
We investigate a simple two-layered viscoelastic rheological model capable of replicating the forced librations of an icy moon with a subsurface ocean. We show that the model, composed only of a prestressed icy crust lying over an effective fluid core (a fixed mantle cavity), can effectively describe the librational behavior of icy moons, thus holding the potential to predict the presence of a subsurface ocean through the analysis of longitudinal librations. The proposed model is applied to the longitudinal librations of Enceladus and Mimas, two small icy moons of Saturn for which relevant data is available from the Cassini mission.

\keywords{Planetary interiors \and Subsurface oceans \and Librations \and Icy moons}
\end{abstract}

%%%%%%%%%%%%%%%%%%%%%%%%%%%%%%%%%%%%%%%%%%%%%%%%%%

%%%%%%%%%%%%%%%%% BODY OF PAPER %%%%%%%%%%%%%%%%%%

\section{Introduction}

The potential of moons and planets to support life depends largely on their internal structure and composition, which makes studying these structures a major goal of modern planetary science. For instance, liquid water is essential for life as we know it, and the discovery of stable reservoirs of liquid water beyond Earth could indicate potentially habitable environments \citep{vance2018}. In our solar system, evidence suggests the presence of subsurface oceans on several icy moons of Jupiter and Saturn, including Titan, Enceladus, Mimas, Europa, Ganymede, and Callisto \citep{nimmo2016}.

Direct \textit{in-situ} geophysical measurements are currently limited to the Earth \citep{Dziewonski1981, tromp2020}, the Moon \citep{Khan2013,matsumoto2015, Garcia2019}, and Mars \citep{khan2021}. For other celestial bodies, one has to rely on alternative indirect approaches to probe their interior composition. Such methods involve analyzing global-scale geophysical data collected by orbiting spacecraft, including mass, mean moment of inertia, tidal response, and body rotational dynamics \citep{bagheri2022}.

%are synchronized with their planets, experiencing small oscillations called librations provoked by the varying gravitational forces as they move through eccentric orbits.

Many moons in our Solar System are in synchronous rotation, where their spin period matches their orbital period. 'Librations' refer to the small forced oscillations of the spin rate around this equilibrium state.

Libration amplitude is a key observable parameter that constrains interior structure, potentially revealing subsurface oceans or decoupled cores \citep{Thomas2016}. Existing detailed models \citep{Hoolst2008,Hoolst2013,Hoolst2016} require extensive interior assumptions.
We present a simplified model to detect subsurface oceans without hypothesizing the full complex interior structure. Limited observational data justifies reduced-parameter models that maintain physical accuracy while enabling efficient analysis \citep{Gev2023}.

Libration amplitudes are usually measured through surface imaging \citep{Thomas2016, Nadezhdina2016, Tajeddine2014} and altimetry \citep{Steinbrugge2019}. The observed librations are those of the external shell. Cassini mission measured the librations for Enceladus \citep{Thomas2016, Nadezhdina2016} and Mimas \citep{Tajeddine2014}; JUICE mission is expected to measure for Ganymede \citep{Steinbrugge2019, Grasset2013, Hussmann2025}. Librations of more distant moons remain unmeasurable with current telescopic capabilities, though future space missions may extend this reach.

\begin{figure*}
\begin{center}
\begin{tikzpicture}[scale=1, transform shape]

\filldraw[gray!50,even odd rule] (0,0) circle (1.8cm) circle (1.5cm);
\filldraw[brown!50,even odd rule] (0,0) circle (0cm) circle (1.2cm);
\filldraw[cyan!30,even odd rule] (0,0) circle (1.2cm) circle (1.5cm);
\draw[black!100, thick] (0,0) circle ( 1.8cm);
\draw[black!100, thick] (0,0) circle ( 1.5cm);
\draw[black!100, thick] (0,0) circle ( 1.2cm);

            \node at (0,-2.2) {Satellite with a global subsurface ocean};

\filldraw[gray!50,even odd rule] (6,0) circle (1.8cm) circle (1.5cm);
\filldraw[brown!50,even odd rule] (6,0) circle (0cm) circle (1.5cm);
\filldraw[cyan!30,even odd rule] (6,0) circle (1.35cm) circle (1.5cm);
\fill[brown!50] (6,1.5) arc (90:-90:1.5) -- cycle;
\draw[black!100, thick] (6,0) circle ( 1.8cm);
\draw[black!100, thick] (6,0) circle ( 1.5cm);

\draw[black!100, thick]  (6,0) +(-90:1.35) arc (-90:-270:1.35);

            \draw [black!100, thick] (6,-1.5) -- (6,-1.35);
            \draw [black!100, thick] (6,1.5) -- (6,1.35);

            \node at (6,-2.2) {Satellite with a partial or no subsurface ocean};

\filldraw[gray!50,even odd rule] (12,0) circle (1.8cm) circle (1.5cm);
\filldraw[cyan!30,even odd rule] (12,0) circle (0cm) circle (1.5cm);
\draw[black!100, thick] (12,0) circle ( 1.8cm);
\draw[black!100, thick] (12,0) circle ( 1.5cm);

            \node at (12,-2.2) {Satellite with a liquid core};

\end{tikzpicture} 
\end{center}
\caption{Three possible internal structures for icy satellites.}\label{fig:int_struc}
\end{figure*}
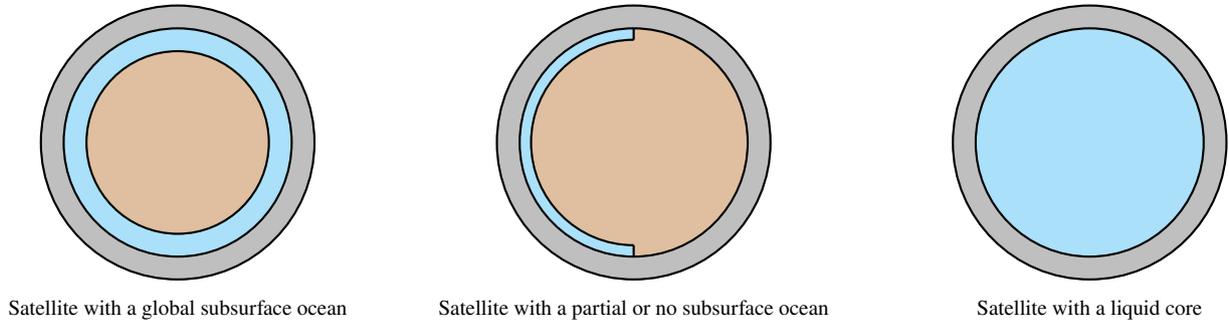

Three interior configurations are possible (Figure \ref{fig:int_struc}): (1) mechanically detached shell over the subsurface ocean, (2) shell at least partially connected with the interior (including rigid satellites), or (3) outer shell overlying a liquid core. Mechanically detached shells require a minimum of two layers to model accurately: a deformable crust overlying an effective liquid layer \citep{RBGR20222}. Bodies without mechanical coupling use effective rheological models based on the total moment of inertia \citep{Tiscareno2009, gev2020}.

We apply a two-layered rheological model in \citep{RBGR20222}, that in its essence is similar to that used by INPOP19a to represent the Moon \citep{Fienga2019}, to study forced longitudinal librations of icy satellites with subsurface oceans. Unlike detailed existing models \citep{Hoolst2008,Hoolst2013,Hoolst2016}, our approach serves as a rapid first test for subsurface ocean detection using few observable parameters.

We detect mechanical core-mantle decoupling by comparing mantle and effective body librations. Note that decoupling can increase or decrease libration amplitude depending on shell thickness and rigidity \citep{Hermes2014}. We apply the model to Enceladus and Mimas, reproducing observations and comparing with literature results. We also calculate energy dissipation rates in the viscoelastic crust and core-mantle boundary.

The paper is organized as follows: Section \ref{sec_dynamical} presents the rheological model, the parameters, and the initial conditions. Section \ref{sec_bodies} applies the model to Enceladus and Mimas with parameter range analysis. Section \ref{results} presents and analyzes the results. Section \ref{conclusions} summarizes the findings and implications.

For clarity, the main symbols and quantities used throughout this paper are summarized in Table \ref{parenc}.

\begin{table}
\begin{center}
\caption{List of symbols in the manuscript}

\begin{tabular}{l l}
\hline
$\kappa$ & Inertial frame at the center of mass of the extended body\\
$G$ & Gravitational constant\\
$M$ & Primary mass\\
$\mathbf{r}$ & Primary position in $\kappa$\\
$\boldsymbol{\omega}_T$, $\boldsymbol{\omega}_m$, $\boldsymbol{\omega}_c \in \kappa$  \hspace{1.0cm} & Tisserand angular velocity of the whole body, mantle, and core \\
$\mat I_m: \kappa\to\kappa$  \hspace{1.0cm} & Moment of inertia of the mantle\\
$\mat I_c: \kappa\to\kappa$  \hspace{1.0cm} & Moment of inertia of the core\\
$\mat I_T = \mat I_m + \mat I_c$ \hspace{1.0cm} & Total Moment of inertia in $\kappa$\\
${\rm I}_{\circ, T}$, ${\rm I}_{\circ, m}$, ${\rm I}_{\circ, c}$ & Mean moment of inertia of the whole body, mantle, and core \\
$\mat B_m: \kappa\to\kappa$ & Deformation operator of the mantle\\
$\mat B_c: \kappa\to\kappa$ & Deformation operator of the core\\
$\mat B_T: \kappa\to\kappa$ & Deformation operator of the whole body\\
$\mat B_0$ & Prestress matrix\\
$\gamma$ &  Gravitational spring rigidity\\
$\mu_0$ & Prestress-elastic constant\\
$\eta_0$ & Prestress-viscosity constant\\
$\eta$ & Viscosity constant of the rheology\\
$k_c$ \hspace{1.0cm} & CMB coupling constant\\
$\boldsymbol{\pi}_T$, $\boldsymbol{\pi}_m$, $\boldsymbol{\pi}_c \in \kappa$ \hspace{1.0cm} & Angular momentum of the whole body, mantle, and core\\
$m_1$ & Mass of the extended body\\
$R_{I}:=\sqrt{\frac{5\rm I_{\circ, T}}{2m_1}}$ & Inertial radius\\
$\nu$ & Kinematic viscosity of the core\\
$R_c$ & Mean radius of the core\\
$n$ & Mean rotation rate\\
$k_2$ &  potential Love number of the extended body\\
$C_{20T}$, $C_{22T}$ & Degree-2 zonal gravity coefficients\\

\hline
\end{tabular} \\[0.3em]
\end{center}
%{\footnotesize  kk}
\label{parenc}
\end{table}

\section{Dynamical model}\label{sec_dynamical}

%% %% %% %% %% %% %% %% %% %% %% %% %% %% %% %% %% %% %% %% %% %% %% %% %% %% %%

%% %% %% %% %% %% %% %% %% %% %% %% %% %% %% %% %% %% %% %% %% %% %% %% %% %% %%

\subsection{Solid-fluid model}\label{sec_solid_fluid}

We consider a two-layer model: an outer viscoelastic, prestressed\footnote{The prestress represents the difference between the body's observed shape and its hydrostatic equilibrium shape without rotation or external fields \citep[Section 4]{RBGR20222}} shell (ice crust) and an effective fluid core (subsurface ocean plus deeper interior) \citep{RBGR20222}. The core-mantle boundary slides as a rigid, spherically symmetric surface over the liquid layer. No further stratification is considered.

This model reproduces forced longitudinal librations of the mantle, directly comparable to observations. We also track the librations of the mean body through its Tisserand frame (body frame with zero total angular momentum). Without mechanical decoupling, Tisserand frame librations match those from non-stratified effective models \citep{gev2020}.

\subsubsection{The equations of motion}\label{sec_eqmot}

We derive the equations of motion for a two-layer deformable body (icy moon) under gravitational influence of a point mass (host planet). Let $\kappa = (\mathbf{i},\mathbf{j},\mathbf{k})$  be an inertial frame at the center of mass of the body. The rotational Lagrangian is \cite{RBGR20222}

\begin{equation}
{{\mathcal{L}}_{\rm{ROT}}} = \frac{1}{2}\boldsymbol{\omega}_m \cdot \mat I_m\boldsymbol{\omega}_m + \frac{1}{2}\boldsymbol{\omega}_c \cdot \mat I_c \boldsymbol{\omega}_c - \frac{3}{2}\frac{GM}{r^5}\mathbf{r} \cdot \mat I_T \mathbf{r} \,,
\end{equation}
where ${\omega}_{\alpha}$  is the angular velocity of layer $\alpha$ (subscripts: $m=\rm{mantle}$, $c = \rm{core}$, $T = \rm{total\,body}$), $\mat I_{\alpha}$ is the corresponding inertia matrix, $M$  is the primary mass, $\mathbf{r}$ is its position, and ${G}=6.6743\times 10^{-11}\mathrm{m}^{3}\cdot\mathrm{kg}^{-1}\cdot\mathrm{s}^{-2}$ is the gravitational constant.

\begin{figure}
\begin{center}
\begin{minipage}{.48\textwidth}
\centering
\begin{tikzpicture}[scale=0.9, transform shape]
\tikzstyle{spring}=[thick,decorate,decoration={zigzag,pre length=0.5cm,post length=0.5cm,segment length=6}]
\tikzstyle{damper}=[thick,decoration={markings,  
  mark connection node=dmp,
  mark=at position 0.5 with 
  {
    \node (dmp) [thick,inner sep=0pt,transform shape,rotate=-90,minimum width=15pt,minimum height=3pt,draw=none] {};
    \draw [thick] ($(dmp.north east)+(5pt,0)$) -- (dmp.south east) -- (dmp.south west) -- ($(dmp.north west)+(5pt,0)$);
    \draw [thick] ($(dmp.north)+(0,-5pt)$) -- ($(dmp.north)+(0,5pt)$);
  }
}, decorate]
\tikzstyle{ground}=[fill,pattern=north east lines,draw=none,minimum width=0.7cm,minimum height=0.3cm]

            \draw [thick] (0,0.7) -- (1,0.7);
            \draw [thick] (1,-0.715) -- (1,2.115);
            \draw [thick] (1.5,2.1) -- (1,2.1);
            \draw [thick] (1.5,-0.7) -- (1,-0.7);
            \draw [damper,thick] (1.5,-0.7) -- (4,-0.7);
            \node at (2.8,-0.2) {$\eta$};
            \draw [thick] (4,-0.7) -- (4.5,-0.7);
            \draw [spring,thick] (1.5,2.1) --  (4,2.1);
            \node at (2.8,2.35) {$\gamma$};
            \draw [spring,thick] (1,0.9) -- (3,0.9);
            \draw [damper,thick] (3,0.9) -- (4,0.9);
            \node at (2,1.15) {$\mu_0$};
            \node at (3.55,1.4) {$\eta_0$};
            \draw [thick] (4,2.1) -- (4.5,2.1);
            \draw [thick] (4.5,-0.715) -- (4.5,2.115);
            \draw [thick] (4,0.9) -- (4.5,0.9);
            \draw [latex-latex, thick] (1,-1.4) -- node[below] {$\varepsilon$} (4.5,-1.4);
            \draw [-latex, thick] (4.5,0.7) -- (5.5,0.7) node[below] {$\sigma$};
            \draw [latex-latex] (1,0.4) -- node[below] {$\varepsilon_0$} (2.8,0.4);
            \draw [latex-latex] (2.8,0.4) -- node[below] {$\tilde\varepsilon_0$} (4.5,0.4);

\node (wall) at (-0.15,0.5) [ground, rotate=-90, minimum width=3cm] {};
\draw [thick] (wall.north east) -- (wall.north west);
      \end{tikzpicture}
\end{minipage}

\end{center}
\caption{Mantle rheological model. The spring $\gamma$
represents the effect of gravity. The damper $\eta$  and the Maxwell element $(\mu_0, \eta_0)$ represent the effect of the macroscopic (spatial average) rheology of the mantle; $\varepsilon$, $\varepsilon_0$ and $\tilde\varepsilon_0$ denote strains and $\sigma$ the stress.
}
\label{fig:simple-osc}
\end{figure}
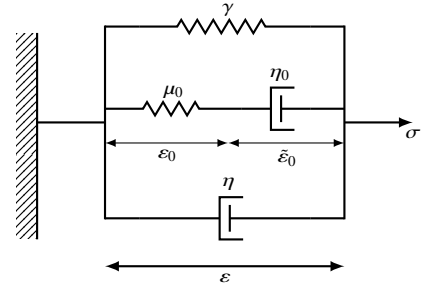

To model deformation of a moon, we assign the macroscopic rheological model (Figure \ref{fig:simple-osc}) to the mantle. The spring with rigidity $\gamma$ represents gravity, the dashpot with viscosity $\eta$ stands for dissipation, and the Maxwell element with elasticity and viscosity $(\mu_0,\eta_0)$ models prestress under the condition $\eta_0\ll\eta$. This is one of the simplest models that considers both viscous and elastic properties of the mantle. If needed, we can easily add any other complex rheology to our model (see \citep{gev2020,Gev2021,Gev2023}), but for the argument of this paper the simple rheology is enough.

Using the Association Principle\footnote{The differential equations for the deformation are derived in \citep{RaR2017} using the Association Principle. The Association principle states: “The differential equations for the deformation $\mat B$ in the body reference frame are equal to the differential equations for the viscoelastic oscillator after replacing $\varepsilon$ by $\mat B$”. The equivalence of the AP and the Correspondence Principle \citep{Efroimsky2012} is addressed in \citep[Section 4]{CRR2018}. The main difference between the Association Principle and the Correspondence Principle is that the first is
formulated in the time domain, while the second is formulated in the frequency
domain.} from \citep{RaR2017}, the deformation Lagrangian is
\begin{equation}
{{\mathcal{L}}_{\rm{TID}}} = \frac{1}{2}\gamma {\rm I}_{\circ,T}\|\mat B_T\|^2 - \frac{1}{2}\mu_0 {\rm I}_{\circ,T}\|\mat B_T -
\mat B_0\|^2\,,
\end{equation}
where $\mat B_{\alpha}$ are deformation matrices satisfying ${\rm I}_{\alpha} = {\rm I}_{\circ, \alpha}(\Id - \mat B_{\alpha})$, with ${\rm I}_{\circ, \alpha} = \displaystyle\frac{1}{3}\rm{Tr}(\mat I_{\alpha})$ the mean moment of inertia and $\mat B_0$ prestress matrix.

The elastic stress within the mantle at the equilibrium state is the prestress $\mu_0(\mat B_T-\mat B_0)$. The prestress allows building a mathematical model to describe the librations of slightly aspherical bodies out of hydrostatic equilibrium and its importance is discussed in \citep{RBGR20222}.
To model the non-conservative forces we have to add the Rayleigh dissipation function to the Lagrangian
\begin{equation}\label{calDdim}
{\cal D} = \frac{1}{2}\eta{\rm I}_{\circ,T}\|\,\dmat B_T-[\boldsymbol{\widehat\omega}_m,\mat B_T]\,\|^2 + \frac{1}{2} k_c\|\boldsymbol{\omega}_m - \boldsymbol{
\omega}_c\|^2\,,
\end{equation}
where $\widehat\omega_{\alpha}$ is an anti-symmetric operator of rotation,\footnote{To every vector $x\in\R^{3}$ we associate an anti-symmetric hat operator defined by
$$x=\begin{pmatrix}
    x_1 \\
    x_2 \\
    x_3 \\
\end{pmatrix}\in\R^{3}\longmapsto\widehat{x}=\begin{bmatrix}
    0 & -x_3 & x_2 \\
    x_3 & 0 & -x_1 \\
    -x_2 & x_1 & 0 \\
\end{bmatrix}\in \mathrm{skew}(3)\,.$$}
and $k_c$ is a core-mantle boundary coupling constant. The total energy dissipation rate is calculated by taking the time-average of the Rayleigh dissipation function, over a full orbital period, once the system has reached a steady state \citep{RaR2017}.

The equations of motion in the inertial frame obtained with Poincaré-Lagrange formalism are \cite{RBGR20222}
\begin{equation}\label{eq:rotdis}
\begin{split}
&\dot{\boldsymbol{\pi}}_m = \mat I_c \boldsymbol{\omega}_c\times\boldsymbol{\omega}_c - k_c(\boldsymbol{\omega}_m - \boldsymbol{\omega}_c) -
3\frac{{G}M}{r^5}(\mat I_T\mathbf{r})\times\mathbf{r} 
\\[0.3em]
&\dot{\boldsymbol{\pi}}_c = \boldsymbol{\omega}_c\times\mat I_c\boldsymbol{\omega}_c + k_c(\boldsymbol{\omega}_m -\boldsymbol{\omega}_c) 
\\[0.3em]
& \eta\dmat B_T = \eta\,[\boldsymbol{\widehat\omega}_m,\mat B_T] - \gamma\mat B_T - \mu_0(\mat B_T-\mat B_0) + \mat F
\\[0.3em]
&\dmat B_c = [\boldsymbol{\widehat\omega}_m, \mat B_c] 
\\[0.3em]
&\dmat B_0 = [\boldsymbol{\widehat\omega}_m, \mat B_0]\,, 
\end{split}
\end{equation}
where
\begin{equation}\label{mod.core}
\begin{split}
&\mat F = -\left(\boldsymbol{\omega}_m\otimes\boldsymbol{\omega}_m-\frac{\omega_m^2}{3}\Id\right) 
+ 3\frac{{G}M}{r^5}\left(\mathbf{r}\otimes\mathbf{r} - \frac{r^2}{3}\Id\right)
\\[0.3em]
&\mat B_m = ({\rm I}_{\circ,T}\mat B_T - {\rm I}_{\circ, c}\mat B_c)/{\rm I}_{\circ, m} 
\\[0.3em]
&\boldsymbol{\omega}_T =  \mat I_T^{-1}\mat I_m\boldsymbol{\omega}_m+
 \mat I_T^{-1}\mat I_c\boldsymbol{\omega}_c=  \boldsymbol{\omega}_m+
  \mat I_T^{-1}\mat I_c\big(\boldsymbol{\omega}_c-\boldsymbol{\omega}_m\big)
\\[0.3em]
&\boldsymbol{\omega}_m = \mat I_m^{-1}\boldsymbol{\pi}_m
\\[0.3em]
&\boldsymbol{\omega}_c = \mat I_c^{-1}\boldsymbol{\pi}_c
\\[0.3em]
&\mat I_T = \mat I_m+\mat I_c\,.
\end{split}
\end{equation}

We complete the two-body system by adding the position and velocity equations for the extended body. To maintain eccentric orbit and achieve forced librations, we fix eccentricity and semi-major axis at present values, neglecting tidal evolution. We also neglect obliquity for simplicity.\footnote{Obliquity can be easily added if needed \citep{RBGR20222}.} The orbital equations of motion are \citep{gev2020}

\begin{equation}\label{eq:position}
\begin{split}
      &  \dot{\mathbf{r}} = \mathbf{v}
        \\[0.3em]
      &  \dot{\mathbf{v}} = - {G}(M+m_1)\frac{1}{r^3}\mathbf{r}\,,
\end{split}
\end{equation}
where $m_1$ is the mass of the extended body.
%Next, we relate model parameters to observable quantities.

%and estimate or assume their values.

%% %% %% %% %% %% %% %% %% %% %% %% %% %% %% %% %% %% %% %% %% %% %% %% %% %% %%
\subsubsection{Model parameters}
%We establish a connection between the parameters of the present model and the ones measured or estimated for the celestial bodies. The rigidity $\gamma$ of the gravitation spring in Figure \ref{fig:simple-osc} is estimated from 
We connect the model parameters to the measured or estimated properties of celestial bodies. The gravitational spring rigidity $\gamma$ (Figure \ref{fig:simple-osc}) is
\begin{equation}
\gamma = \frac{4}{5}\frac{{G}m_1}{R_{I}^3}\,, \qquad R_I = \sqrt{\frac{5\rm I_{\circ, T}}{2m_1}}\,,
\end{equation}
where $R_I$ is the inertial radius \cite{ragazzo2020theory}. The core-mantle boundary (CMB) coupling constant, responsible for the boundary dissipation, cannot be observationally constrained except in rare cases like the Moon. We estimate it using the theoretical approach of \citep{peale2014} for Mercury, which employs fluid dynamic arguments to derive
\begin{equation}
k_c = \nu\frac{1}{R_c^2}\frac{\rm I_{\circ, c}\rm I_{\circ, m}}{\rm I_{\circ, T}}\,,   
\end{equation}
where $R_c$ is the mean core radius and $\nu$ is the kinematic viscosity of the core fluid. In \citep[section 3.4]{RBGR20222} authors detail how to derive $k_c$ from CMB and core fluid physical characteristics, and justify replacing the kinematic viscosity with the eddy viscosity.
The core-mantle coupling constant $k_c$ represents momentum transfer across the liquid interface. While derived for Mercury's metallic core, the functional form $k_c \propto \nu / R_c^2$ remains valid for icy moon oceans, though the effective viscosity $\nu$ may differ significantly from Mercury's core.

The mean moments of inertia for the mantle, core, and total body relate to the gravitational coefficients of the satellite as
\begin{equation}
\begin{split}
&\frac{{\rm I}_{\circ, T}}{MR_T^2} = \frac{C_T}{MR^2_T} +
\frac{2}{3}C_{20T}
\\[0.3em]
&\frac{{\rm I}_{\circ, c}}{MR_T^2} = \frac{C_c}{MR_T^2} +
\frac{2}{3}C_{20c}
\\[0.3em]
&\frac{{\rm I}_{\circ, m}}{MR_T^2} = 
  \frac{{\rm I}_{\circ, T}}{MR_T^2}
- \frac{{\rm I}_{\circ, c}}{MR_T^2}\,,
\end{split}
\end{equation}
where $C_{20\alpha}$ are the unitless degree-2 zonal gravity coefficients describing the gravity field and $\frac{C_{\alpha}}{MR^2_T}$ are polar moments of inertia.

The prestress elastic constant depends on the potential Love number and mean moment of inertia, and does not depend on mantle viscosity \citep[section 10.1]{RBGR20222}:
\begin{equation}
    \mu_{0}  = 3\frac{{G}(M + m_{1})}{1+M/m_{1}}\frac{{\rm I}_{\circ, T}}{MR_T^2}\frac{1}{R_{T}^{3}k_{2}}
\end{equation}
where $k_2$ is the potential Love number of the deformable body.

\subsection{Initial conditions and integration of the model}\label{sec:initial_cond}
We specify the initial conditions for some key variables needed to integrate the equations of motion. The prestress matrix requires accurate theoretical or experimental values, since it remains constant during integration and critically affects longitudinal librations. This matrix is evaluated from the mean moments of inertia of the extended body as
\begin{equation}\label{prestress}
    \begin{split}
&        \mat B_0(0) = \frac{2}{3}\frac{MR_T^2}{{\rm I}_{\circ,T}}
  \begin{bmatrix}
  3C_{22T}-\frac{1}{2}C_{20T} & 0 & 0 \\
  0 & -3C_{22T}-\frac{1}{2}C_{20T} & 0 \\
  0 & 0 & C_{20T}
  \end{bmatrix}
\\[0.3em]
& + \frac{n^2}{3\mu_0}
  \begin{bmatrix}
  -1 & 0 & 0 \\
  0 & -1 & 0 \\
  0 & 0 & 2
  \end{bmatrix}\,.
    \end{split}
\end{equation}
The first matrix is the gravitational potential variation from the observed shape, and the second is the rotational effect in the prestressed state, where $n^2$ is the squared mean rotation rate. The second term reflects the coupling between orbital motion and internal stress distribution.
The initial deformation of the core is estimated from its moment of inertia
\begin{equation}
    \begin{split}
\mat B_c(0) = \frac{2}{3}\frac{MR_T^2}{{\rm I}_{\circ, c}}
  \begin{bmatrix}
  -\frac{1}{2}C_{20c} & 0 & 0 \\
  0 & -\frac{1}{2}C_{20c} & 0 \\
  0 & 0 & C_{20c}
  \end{bmatrix}\,.
    \end{split}
\end{equation}

Orbital and rotational parameters are set to current values. The total body deformation matrix is initialized as in \citep{RaR2017,gev2020}.

The system of first-order ODEs (\ref{eq:rotdis}, \ref{eq:position}) is integrated using an 8th-order Runge-Kutta method with 7th-order error estimation (Dormand-Prince) and adaptive step size \citep{HairerEDO1993}. To obtain the results in Section \ref{results}, the system was integrated for a duration of 500 orbital periods to ensure that any initial free librations were fully damped by the dissipative terms leaving only the steady-state forced libration response. The initial conditions were set to the moon's present-day orbital and rotational parameters.

%% %% %% %% %% %% %% %% %% %% %% %% %% %% %% %% %% %% %% %% %% %% %% %% %% %% %%

\section{Applications: Enceladus and Mimas}\label{sec_bodies}

We test our model on Enceladus and Mimas, two Saturn moons of similar size on neighboring orbits but with different surface patterns suggesting distinct internal structures and evolutionary histories. Cassini mission measured the forced longitudinal libration amplitudes for both Enceladus \citep{Thomas2016, Nadezhdina2016} and Mimas \citep{Tajeddine2014}. The subsurface ocean in Enceladus is supported by the saltiness of its south polar plumes; longitudinal libration amplitude makes the ocean global \citep{Thomas2016}. Mimas's anomalously large libration amplitude was long interpreted as indicating either a non-hydrostatic core or a global subsurface ocean \citep{Tajeddine2014}. A recent study has since confirmed the presence of a young ocean, making Mimas a key target for understanding ocean world evolution \citep{Lainey2024}. We model both moons with global subsurface oceans (Figure \ref{fig:int_struc}, left panel) using our two-layer approach: icy crust over an effective liquid core. Saturn is treated as a point mass with $M=5.683\times 10^{26}\,\mathrm{kg}$ (NASA Planetary Fact Sheets). The physical and orbital parameters are given in Table \ref{tab:parameter}. We explore ranges of shell thickness, rigidity, and mantle viscosity to optimize agreement with observed libration amplitudes.

\begin{table}
\begin{center}
\caption{Present day physical and orbital parameters of Enceladus and Mimas}
\label{tab:parameter}
\begin{tabular}{l l l }
\hline

 \hspace{0.7cm} & Enceladus \hspace{0.7cm} & Mimas  \\
\hline
$m_{1}$  & $1.08\times 10^{20}\,\mathrm{kg}$ & $0.379\times 10^{20}\,\mathrm{kg}$  \\

$R$   & $252.1\times 10^{3}\,\mathrm{m}$  &  $198.66\times 10^{3}\,\mathrm{m}$ \\

$e$   & $0.0045$  & $0.0202$  \\

$a$   & $2.38\times 10^{8}\,\mathrm{m}$  & $1.8552\times 10^{8}\,\mathrm{m}$  \\

$n$   & $5.308\times 10^{-5}\,\mathrm{rad}\cdot\mathrm{s}^{-1}$  & $7.716\times 10^{-5}\,\mathrm{rad}\cdot\mathrm{s}^{-1}$  \\

${\rm I}_{\circ, T}$         &  $2.2994\times 10^{30}\,\mathrm{kg}\cdot\mathrm{m}^{2}$  & $5.57187\times10^{29}\,\mathrm{kg}\cdot\mathrm{m}^{2}$  \\

           $I_{oT}/MR^2_T$ & $0.335$ & $0.375$ \\
 $C_T/MR^2_T$ & $0.3386$ & $0.385$ \\
  $C_{20T}$ & $-5.4352\times10^{-3}$ & $-1.5\times10^{-2}$ \\
 $C_{22T}$ & $1.5498\times10^{-3}$ & $3.6\times10^{-3}$ \\
 
$\gamma$            &  $4.69504\times 10^{-7}\,\mathrm{s}^{-2}$  &  $7.33073\times 10^{-7}\,\mathrm{s}^{-2}$ \\
          $\nu$ & $0.3\,\mathrm{m^2}\cdot\mathrm{s}^{-1}$ & $0.3\,\mathrm{m^2}\cdot\mathrm{s}^{-1}$\\
%$\eta$ & $10^{18} \mathrm{Pa}\cdot\mathrm{s}$\\

$k_{2}$            & $0.0167^{+0.0403}_{-0.0119}$  &  $0.01$ \\
\hline
\end{tabular} \\[0.3em]
\end{center}
{\footnotesize  Masses, radii, eccentricities, semi-major axis, and rotation rate of both moons are taken from NASA Planetary Fact Sheets. The rigidity constant $\gamma$ is calculated for both bodies using the method exposed in \citep{CRR2018}. The gravitational coefficients $C_{20T}, C_{22T}$ and the total moment of inertia of Enceladus were taken from \citep{Iess2014}. The mean moment of inertia of Mimas is taken from \citep{Szilard2022}, and the gravitational coefficients are estimated. An estimate of the eddy viscosity $\nu$ of the fluid inside the core of Enceladus is taken from  \citep{Kang2022}; the same eddy viscosity is used for the ocean of Mimas. The potential Love number $k_{2}$ of Enceladus is taken from \citep{Ermakov2021}, and that of Mimas from \citep{Lainey2024}.}
\end{table}

\subsection{Enceladus interior model}

Enceladus maintains a synchronous, slightly eccentric orbit around Saturn through a 2:1 mean motion resonance with Dione. Analysis of Cassini spacecraft data has yielded two independent measurements of its libration amplitude: $0.120^{\circ} \pm 0.014^{\circ}$ \citep{Thomas2016} and $0.155^{\circ} \pm 0.014^{\circ}$ \citep{Nadezhdina2016}.

We consider the following parameter set for the interior of Enceladus. We fix the density of the ice crust at $850 \,\mathrm{kg/m^{3}}$. This density is different from the standard value for pure ice ( $\sim917 \,\mathrm{kg/m^{3}}$) and is used, for example, in \citep{Thomas2016} as a reasonable assumption for a slightly porous ice shell. Shell thickness varies from $2$ to $60\,\mathrm{km}$ to assess the dependence of the libration amplitude and identify the optimal fit values. Three prestress elasticity constants are tested: $0.27\,\mathrm{GPa}$, $0.54\,\mathrm{GPa}$, and $1.08\,\mathrm{GPa}$, corresponding to the potential Love numbers $0.0055$, $0.011$, and $0.022$, respectively, within the uncertainty range of \citep{Ermakov2021}. Ice shell viscosity spans $10^{12}$--$10^{17}\, \mathrm{Pa\cdot s}$ \citep{Behounkova2015, efr2018, Hoolst2016, gev2020}. The core moments of inertia scale linearly with the size of the core.

\subsection{Mimas interior model}

Mimas is on a synchronous orbit around Saturn with significantly higher eccentricity than Enceladus, though the origin of this anomalously large present-day eccentricity remains unexplained. Cassini observations yield a longitudinal libration amplitude of $0.8383^{\circ}\pm0.017^{\circ}$ \citep{Tajeddine2014}.

Several factors complicate libration modeling for Mimas. First, its shell's oblate shape significantly influences its forced longitudinal librations \citep{Lainey2024}. Second, Mimas approaches the lower limit of size and mass for achieving hydrostatic equilibrium, making prestress effects potentially crucial for accurate libration predictions. In our model, prestress is calculated using Equation (\ref{prestress}) and depends primarily on the first-order gravitational coefficients $J_{2}$ and $C_{22}$.

Unlike Enceladus, no observational constraints exist for Mimas's gravitational coefficients. We therefore treat these as free parameters, adjusting them to match the observed libration amplitude while using the estimates from \citep{Lainey2024} as reference values. The observed libration amplitude of Mimas ($0.838^{\circ}$) cannot be reproduced using gravitational coefficients from \citep{Lainey2024} without unrealistic ice shell thicknesses ($>34 \rm{km}$). This discrepancy likely reflects: (1) uncertainties in Mimas' gravitational field due to limited observational constraints, (2) possible non-hydrostatic effects not captured in simple models, or (3) additional interior complexities. We adjust $J_{2}$ and $C_{22}$ by $~20\%$ from \citep{Lainey2024} to achieve observed librations with reasonable shell thickness ($28 \rm{km}$). This adjustment is within expected uncertainties for Mimas' poorly constrained gravitational field and suggests future missions should prioritize precise gravity measurements.

Following our Enceladus analysis, we assume an identical crustal density of $850 \,\mathrm{kg/m^{3}}$ and explore ice shell thicknesses ranging from $6$ to $80\,\mathrm{km}$. Multiple combinations of potential Love number and crustal viscosity are tested to reproduce the observed libration amplitude.

\section{Results}\label{results}
\subsection{Libration amplitude}

We analyze how mantle size affects the longitudinal libration amplitude of both the mantle and Tisserand frame. Using potential Love number $k_{2} = 0.011$ and mantle viscosity $\eta = 10^{14}\,\mathrm{Pa\cdot s}$ for Enceladus, we obtain libration dependence on ice shell thickness.

The forced longitudinal libration amplitude does not appear explicitly in the equations of motion (\ref{eq:rotdis}, \ref{eq:position}), and once the free
libration is totally damped, is given by \cite{gev2020}
\begin{equation}
    \beta = \sqrt{\bigg(\frac{\omega_3}{n}-1\bigg)^2+\bigg(\frac{\dot{\omega}_3}{n^2}\bigg)^2},
\end{equation}
where $n$ is the orbital frequency and $\omega_3$ is the third component of the total or mantle angular velocity. The two quantities inside the square root are obtained by numerical integration of the equations of motion (\ref{eq:rotdis}, \ref{eq:position}).

\begin{figure}
\centering
\includegraphics[scale=0.65]{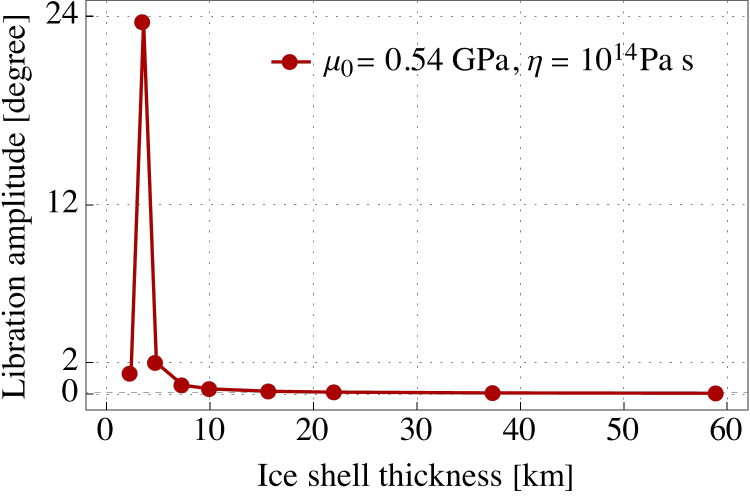}\\

\caption{Amplitude of Enceladus's longitudinal libration (in degrees) as a function of ice shell thickness.\label{fig:enc_res}}
\end{figure}

We observe a very large libration amplitude for a $3.6\,\rm{km}$ ice crust (see Figure \ref{fig:enc_res}), consistent with resonant behavior in \citep[Figure 2]{Hoolst2016}. This resonant behavior is expected for subsurface ocean models, where the ocean creates a free libration frequency that can resonate with orbital frequency. This frequency depends on shell thickness \citep{Rambaux2011}.
The spontaneous appearance of the resonance demonstrates that our two-layer model correctly reproduces subsurface ocean signatures. Outside the resonance, libration amplitude decreases with increasing crust thickness, as expected.

\begin{figure}
\centering
\includegraphics[scale=0.65]{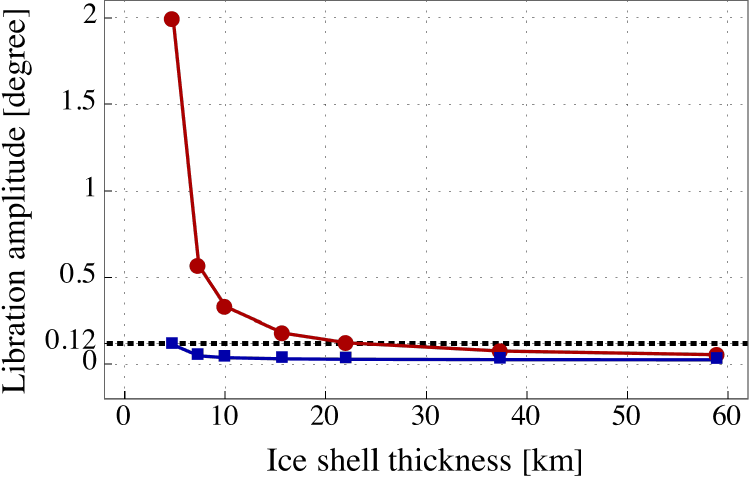}\\

\caption{Amplitude of Enceladus's longitudinal libration (in degrees) as a function of ice shell thickness. The red line with dots stands for the shell libration and the blue line with squares for the librations of bodies Tisserand frame. The thin hatched (line-like) interval shows the measured libration amplitude with the uncertainty.
\label{fig:crust_tis}}
\end{figure}

In Figure \ref{fig:crust_tis} we compare the mantle and Tisserand frame librations. The mean body libration amplitude is much smaller than the mantle libration, matching observations only for unrealistically thin ice crusts. This confirms that homogeneous models underestimate libration amplitudes for bodies with subsurface oceans \citep{Hoolst2016, gev2020}. We obtain the libration amplitude estimate in \citep{Thomas2016} with potential Love number $k_2\approx0.011$, ice shell viscosity $\eta\approx10^{14}\, \mathrm{Pa\cdot s}$, and shell thickness $\approx22\, \mathrm{km}$, consistent with \citep{Hoolst2016} (see Figure \ref{fig:crust_tis}). This viscosity is typical for thin ice layers at the melting point \citep{Soucek2019}, though bulk ice viscosity should be higher to avoid overestimating crustal dissipation \citep{Rhoden2022}.

\begin{figure}
\centering
\includegraphics[scale=0.65]{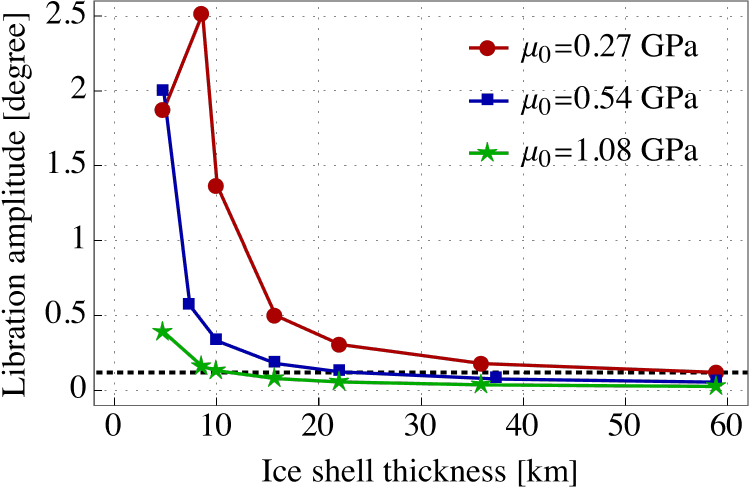}\\

\caption{Amplitude of Enceladus's longitudinal libration (in degrees) as a function of ice shell thickness for three values of the prestress-elastic constant. The thin hatched (line-like) interval shows the measured libration amplitude with the uncertainty.\label{fig:elas_var}}
\end{figure}

The libration amplitude decreases with the increase of the prestress-elastic constant (see Figure \ref{fig:elas_var}), which is opposite to the libration dependence on mantle rigidity \citep{Hermes2014, Hoolst2016}. These parameters should not be confused: rigidity controls elastic behavior while the prestress-elastic constant determines how much the mantle is prestressed.

\begin{figure}
\centering
\includegraphics[scale=0.65]{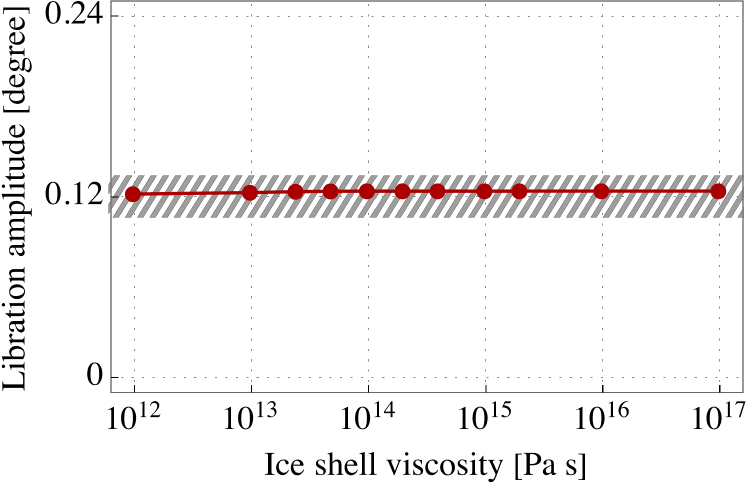}\\

\caption{Amplitude of Enceladus's longitudinal libration (in degrees) as a function of ice shell viscosity for an ice shell of $\sim 22$ km. The hatched interval shows the measured libration amplitude with the uncertainty.\label{fig:enc_vis_var}}
\end{figure}

Viscosity variations over 5 orders of magnitude ($10^{12}-10^{17}\, \mathrm{Pa\cdot s}$) do not significantly affect the libration amplitude (Figure \ref{fig:enc_vis_var}), indicating that this parameter cannot be uniquely constrained from libration data alone, which agrees with \citep{Hoolst2016} for the $10^{13}-10^{15}\, \mathrm{Pa\cdot s}$ range. This degeneracy limits our model's ability to independently constrain interior viscosity structure and requires additional observational constraints (e.g., thermal models, tidal heating estimates) for complete characterization.

\begin{figure}
\centering
\includegraphics[scale=0.65]{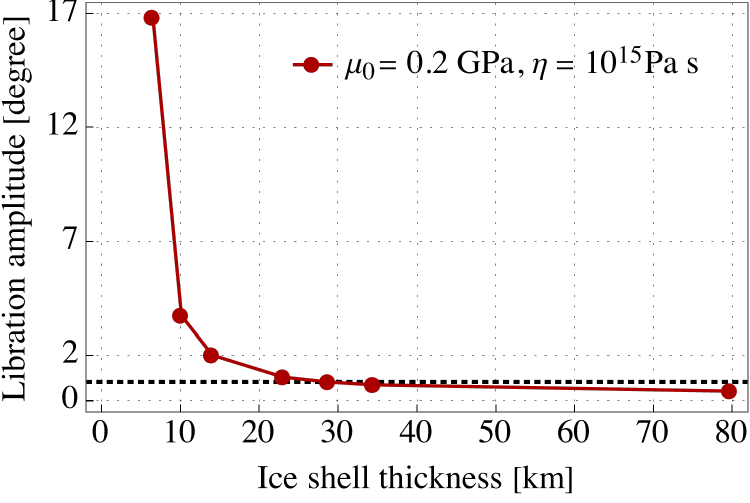}

\caption{Amplitude of Mimas's longitudinal libration (in degrees) as a function of ice shell thickness. The thin hatched (line-like) interval shows the measured libration amplitude with the uncertainty.\label{fig:mim_lib_rad}}
\end{figure}

\begin{figure}
\centering
\includegraphics[scale=0.65]{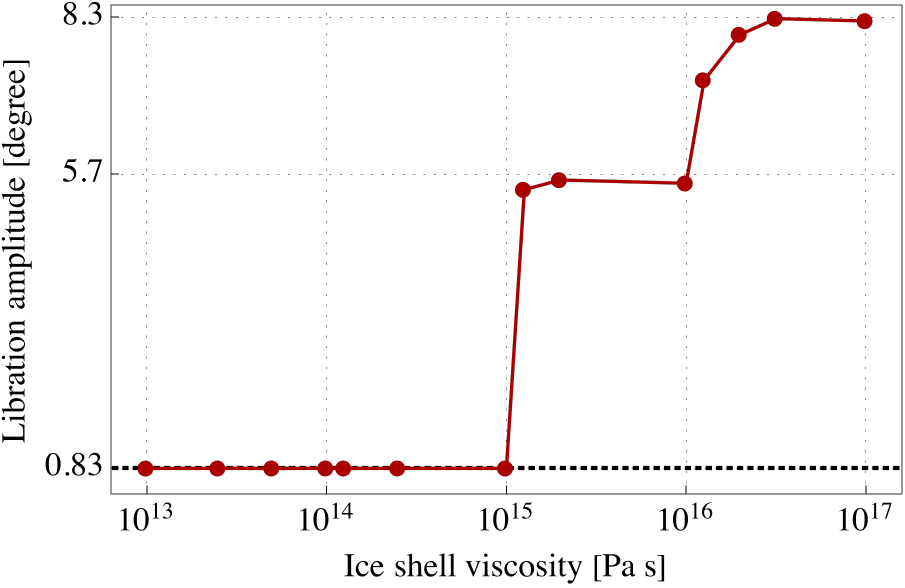}\\

\caption{Amplitude of Mimas's longitudinal libration (in degrees) as a function of ice shell viscosity for an ice shell of $\sim 28$ km. The thin hatched (line-like) interval shows the measured libration amplitude with the uncertainty.\label{fig:mim_vis_var}}
\end{figure}

Mimas's prestress dominates the forced librations of the crust, with viscosity and potential Love number having minimal effect on amplitude. We use the libration amplitude to constrain first-order gravitational coefficients. Initially using $J_{2} = 0.01875$ and $C_{22} = 0.0045$, 
$k_{2} = 0.01$ (equivalent to $\mu_{0}=2\times 10^{8}\, \mathrm{Pa}$), and $\eta=10^{15}\, \mathrm{Pa\cdot s}$ \citep{Lainey2024}, the observed libration requires $\sim 34.5\mathrm{km}$ shell thickness, outside the range constrained by \citep{Tajeddine2014}. Reducing shell prestress to $\sim80\,\%$ of the \citep{Lainey2024} value while maintaining Love number and viscosity yields reasonable shell thickness. Solving Equation (\ref{prestress}) gives revised gravitational coefficients (Table \ref{tab:parameters}), that reproduce the observed libration with $\sim28\,\mathrm{km}$ crust thickness (Figure \ref{fig:mim_lib_rad}), consistent with \citep{Tajeddine2014}. This thickness remains constant for Love numbers $0.00057-0.01$ and viscosities $10^{13}-10^{15}\, \mathrm{Pa\cdot s}$. Above $10^{15}\, \mathrm{Pa\cdot s}$, libration amplitude increases with viscosity as expected \citep{Hermes2014} (Figure \ref{fig:mim_vis_var}).

The required parameter adjustment for Mimas (reducing prestress by $~20\%$) may indicate either: (a) systematic errors in current gravitational coefficient estimates, or (b) limitations of our simplified model for this particular moon. Future Cassini data reanalysis or new missions could resolve this ambiguity.

\subsection{Dissipation}\label{dissipation}

\begin{figure}
\centering
\includegraphics[scale=0.65]{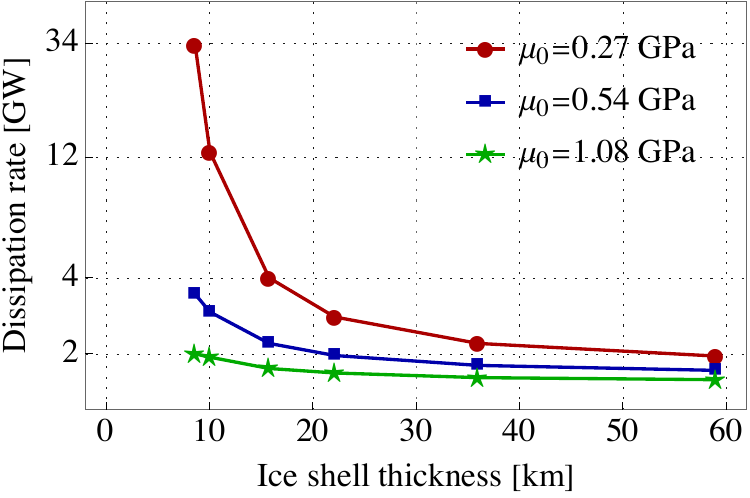}\\

\caption{Enceladus's dissipation as a function of ice shell thickness.\label{fig:enc_dis}}
\end{figure}

\begin{figure}
\centering
\includegraphics[scale=0.65]{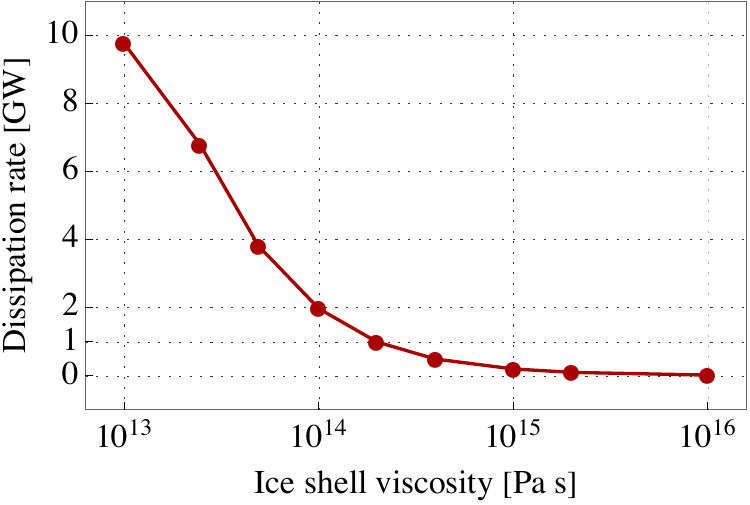}\\

\caption{Enceladus's dissipation as a function of ice shell viscosity with $\mu_{0} = 0.54 \mathrm{GPa}$ and ice shell thickness of $22\, \mathrm{km}$.\label{fig:enc_vis_dis}}
\end{figure}

\begin{figure}
\centering
\includegraphics[scale=0.65]{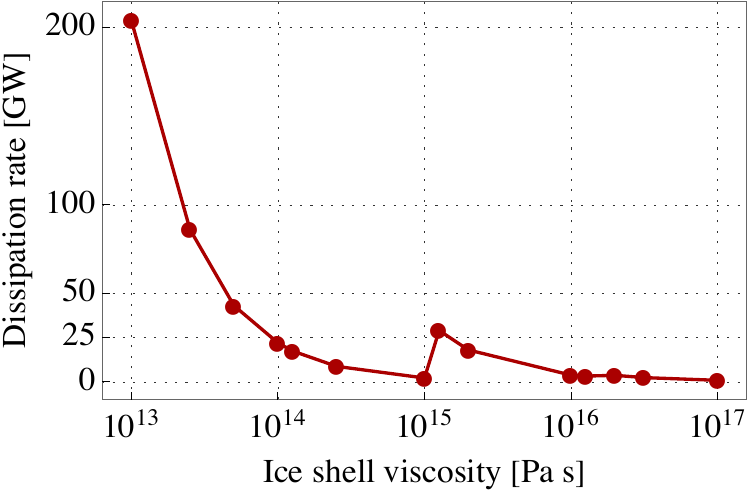}\\

\caption{Mimas's dissipation as a function of ice shell viscosity with $\mu_{0} = 0.2 \mathrm{GPa}$ and ice shell of $28\mathrm{km}$.\label{fig:mim_vis_dis}}
\end{figure}

We estimate energy dissipation in the mantle and at the core-mantle boundary for both Enceladus and Mimas. The energy is decreasing along the motion \cite{RaR2017,RBGR20222}:
\begin{equation}
    \dot{E} = -2\mathcal{D},
\end{equation}
where $\mathcal{D}=\mathcal{D}_{m} + \mathcal{D}_{CMB}$ is given by equation (\ref{calDdim}). CMB dissipation is orders of magnitude smaller than mantle dissipation. Since our model neglects core dissipation, the rates in Figures \ref{fig:enc_dis}, \ref{fig:enc_vis_dis}, and \ref{fig:mim_vis_dis} do not represent the total energy balance.

For Enceladus, the mantle dissipation decreases with increasing $\mu_{0}$ (decreasing potential Love number) (see Figure \ref{fig:enc_dis}), mirroring mantle libration behavior. Once librations are fixed, the Love number is also constrained, making shell viscosity the primary control on dissipation variation for fixed libration amplitude. Figure \ref{fig:enc_vis_dis} confirms that dissipation decreases with increasing ice viscosity, consistent with the other models in the literature \citep{Tobie2005}.
For $\eta = 10^{14}\, \mathrm{Pa\cdot s}$, Enceladus mantle dissipation reaches $\sim2.2\,\mathrm{GW}$. This is an upper bound for Maxwell rheology without resonances. Alternative rheologies (Burgers, Andrade, Sundberg-Cooper) can be considered to possibly increase the dissipation in the mantle \citep{gev2020,Gev2021,Gev2023}. This value is lower than estimates from Cassini data,  $5.8 \pm 1.9$ GW \citep{Spencer2006}, $15.8 \pm 3.1$ GW \citep{Howett2011}, or $3.9 - 18.9$ GW from a theoretical study in \citep{Hay2019}. This discrepancy is expected, as our model only accounts for dissipation within the viscoelastic ice shell and on the shell ocean boundary. The higher observed values likely include significant additional dissipation from tides within the ocean itself or at the seafloor, or in the porous core \citep{Choblet2017}, which are not included in our model.

Figure \ref{fig:mim_vis_dis} shows Mimas shell and CMB dissipation versus viscosity. Dissipation decreases with increasing viscosity during almost constant libration plateaus (Figure \ref{fig:mim_vis_var}) but increases abruptly between plateaus. For a $28\,\mathrm{km}$ shell with $\mu_{0} = 0.2 \mathrm{GPa}$ and $\eta = 10^{15}\, \mathrm{Pa\cdot s}$, mantle and CMB dissipation reaches $\sim2.17\,\mathrm{GW}$.

Our model predicts similar tidal dissipation rates ($\sim2.2\,\mathrm{GW}$) for both Enceladus and Mimas despite their different orbital eccentricities, suggesting our simplified treatment may miss important distinguishing physics. For subsurface ocean maintenance, additional heating sources are required: radiogenic heating ($\sim0.1-0.3\,\mathrm{GW}$ from chondritic composition), orbital evolution (potentially significant but model-dependent), and core dynamics (not quantified here).

%% %% %% %% %% %% %% %% %% %% %% %% %% %% %% %% %% %% %% %% %% %% %% %% %% %% %%

\section{Conclusions}\label{conclusions}

Our two-layer rheological model effectively approximates icy moons with global subsurface oceans for forced longitudinal libration modeling. The model is simple, easily implemented, and preserves key physical behaviors of ocean-bearing bodies, making it valuable for preliminary internal structure analysis using libration data. This simplified approach enables rapid assessment of newly discovered moons and initial mission data analysis, detecting subsurface oceans and estimating crustal thickness from libration amplitudes. The approach is best suited for preliminary analysis and mission planning rather than detailed interior characterization. The model extends to latitudinal librations \citep{RBGR20222} and realistically predicts resonances in dissipation and libration amplitude expected for global subsurface ocean bodies \citep{Rambaux2011}.

Our two-layer model cannot distinguish a subsurface ocean from a deep fluid core; it only demonstrates that any large, mechanically decoupled liquid layer, be it a subsurface ocean or a fluid core, will produce a significantly larger libration than a solid body. Distinguishing between these two scenarios would require additional geophysical constraints, such as the tidal Love number $k_2$ or detailed geological mapping, which is beyond the scope of this simplified dynamical model.

Application to Enceladus and Mimas demonstrates how libration amplitude depends on crustal thickness, viscosity, and body prestress. Shell thickness estimates ($\sim22 \rm{km}$ for Enceladus, $\sim28 \rm{km}$ for Mimas) agree with literature values from complex models \citep{Hoolst2016, Tajeddine2014}. 

We emphasize that our simplified two-layer model has limitations. It focuses solely on the primary orbital forcing frequency and neglects mutual gravitational perturbations from other satellites, assumes uniform viscosity in the shell, and does not capture dissipation within the ocean and core, lacking energy sources needed to balance the moons' energy budgets. Additionally, it neglects higher-order orbital perturbations that could introduce resonant effects. Significant parameter degeneracy prevents independent constraints on viscosity and Love number from libration amplitude alone, limiting the model's diagnostic capabilities. The model is therefore best suited for preliminary analysis and mission planning rather than detailed interior characterization, for which more comprehensive multi-layer models are required.

\begin{acknowledgements}

The author thanks Clodoaldo Ragazzo (IME/USP) for valuable discussions during the preparation of the manuscript, and Diogo A. Gomes (KAUST) for kind and fruitful encouragement. The author gratefully acknowledges generous financial support from KAUST, Saudi Arabia, and partial funding from FAPESP, Brazil, through grant 2019/25356-9.
\end{acknowledgements}


\bibliographystyle{abbrv}
\bibliography{mybibliography} % if your bibtex file is called mybibliography.bib

\end{document}